\documentclass[english,aps,prc,groupedaddress,amsmath,amssymb,showpacs]{revtex4}
\usepackage[T1]{fontenc}
\usepackage[latin9]{inputenc}

\makeatletter

\providecommand{\tabularnewline}{\\}

\@ifundefined{textcolor}{}
{%
 \definecolor{BLACK}{gray}{0}
 \definecolor{WHITE}{gray}{1}
 \definecolor{RED}{rgb}{1,0,0}
 \definecolor{GREEN}{rgb}{0,1,0}
 \definecolor{BLUE}{rgb}{0,0,1}
 \definecolor{CYAN}{cmyk}{1,0,0,0}
 \definecolor{MAGENTA}{cmyk}{0,1,0,0}
 \definecolor{YELLOW}{cmyk}{0,0,1,0}
 }

\usepackage{babel}\usepackage{longtable}

\makeatother

\usepackage{babel}
\begin{document}

\title{How degeneracies can obscure interesting physics}

\author{L. Zamick}

\author{A. Escuderos}

\affiliation{Department of Physics and Astronomy, Rutgers University, Piscataway,
New Jersey, 08854 USA}
\begin{abstract}
We show how degeneracies, accidental or otherwise, can obscure some
interesting physics. We further show how one can get around this problem. 
\end{abstract}
\maketitle

In a 2006 publication~\cite{ez06} Escuderos and Zamick found some
interesting behaviour in the $g_{9/2}$ shell. Unlike the lower shells,
e.g. $f_{7/2}$, seniority is not a good quantum number in the $g_{9/2}$
shell. Despite this, it was found that in a matrix diagonalization
with four identical particles in the $g_{9/2}$ shell with total angular
momentum $I=4$ or 6, one unique state emerged no matter what interaction
was used. Before the mixing one has two states with seniority $v=4$
and one with $v=2$. The surprise was that, after the diagonalization,
one gets a unique state that is always the same independently of the
interaction used. This unique state has seniority $v=4$. The components
of the wave function are given in the third column after $[J_{p}\,,J_{n}]$.
The problem to be dealt with was not only why this state did not mix
with the $v=2$ state but also why it does not mix with the other
$v=4$ state. But this will not concern us here. Rather we will use
this as an example of how degeneracies can obscure interesting physics.

It was already commented on in the 2007 paper~\cite{z07} that cfp's
for identical particles are usually calculated using a pairing interaction.
With such an interaction, the two $v=4$ states are degenerate, i.e.
they have the same energy. This means that any linear combination
of the two states can emerge in a matrix diagonalization. One can
get different combinations with different programs or even with the
same program run at different times. Thus the emergence of a unique
state gets completely lost.

In this work we consider a less obvious example: a matrix diagonalization
of two proton holes and two neutron holes in the $g_{9/2}$ shell,
i.e. we consider $^{96}$Cd rather than $^{96}$Pd, the latter consisting
of four proton holes (whether we consider holes or particles does
not matter). We use a quadrupole--quadrupole interaction $Q\cdot Q$
for the matrix diagonalization. The two-body matrix elements in units
of MeV from $J=0$ to $J=9$ are: $-1.0000$, $-0.8788$, $-0.6516$,
$-0.3465$, $-0.0152$, 0.2879, 0.4849, 0.4849, 0.1818, and $-0.5454$.

We show the results in Table~\ref{tab:res}. For $I=4$ we get 14
eigenfunctions, but we list only two of them in the first two columns.
The reason we single these out is that they are degenerate---both
are at an exciation energy of 3.5284~MeV.

In the third wave function column we have the unique state, one that
emerges, as we said above, with any interaction, however complicated,
e.g. CCGI~\cite{ccgi12}. But now we have to modify the phrase {}``any
interaction''. We do not see this unique state when we use the $Q\cdot Q$
interaction---none of the 14 states looks like the one in column 3.
Learning from our experience with the pairing interaction, we suspect
that the problem lies with the two degenerate states at 3.5284~MeV.
We assumed that the two states were mixtures of one $T=0$ and one
$T=2$ state.

\begin{table}[htb]
 \caption{\label{tab:res} Selected $I=4^{+}$ states in $^{96}$Cd with a $(g_{9/2})^{4}$
configuration. On the second row we give the energies in MeV.}

\begin{ruledtabular} %
\begin{tabular}{cccccc}
 & Mix $T=0,2$  & Mix $T=0,2$  & $T=2$, $v=4$ unique  & $T=0$ untangled  & other $T=2$, $v=4$ \tabularnewline
 & 3.5284  & 3.5284  & 6.5285  & 3.5284  & \tabularnewline
$[J_{p}\,,J_{n}]$  &  &  &  &  & \tabularnewline
$[0\,,4]$  & 0.0000  & 0.0000  & 0.0000  & 0.0000  & 0.0000 \tabularnewline
$[2\,,2]$  & $-0.3250$  & $-0.4170$  & $-0.4270$  & 0.3123  & $-0.0255$ \tabularnewline
$[2\,,4]$  & $-0.2364$  & $-0.2472$  & $-0.2542$  & 0.2289  & $-0.1986$ \tabularnewline
$[2\,,6]$  & 0.2168  & 0.3043  & 0.3107  & $-0.2076$  & $-0.1976$ \tabularnewline
$[4\,,4]$  & 0.0207  & 0.2390  & 0.2395  & $-0.0135$  & $-0.3313$ \tabularnewline
$[4\,,6]$  & $-0.1826$  & $-0.1364$  & $-0.1418$  & 0.1784  & 0.2245 \tabularnewline
$[4\,,8]$  & 0.0934  & 0.1540  & 0.1567  & $-0.0888$  & 0.3874 \tabularnewline
$[6\,,6]$  & $-0.1312$  & 0.1678  & 0.1638  & 0.1362  & 0.5645 \tabularnewline
$[6\,,8]$  & $-0.1343$  & 0.0357  & 0.0316  & 0.1353  & 0.0247 \tabularnewline
$[8\,,8]$  & $-0.7421$  & 0.5881  & 0.5625  & 0.7594  & $-0.1087$ \tabularnewline
\end{tabular}\end{ruledtabular} 
\end{table}

We can remove the degeneracy without altering the wave functions of
the non-degenerate states by adding a $t(1)\cdot t(2)$ interaction
to the Hamiltonian. This will shift energies of states of different
isospin. What we actually did was equivalent to this. We added $-1.000$
MeV to the two-body $T=0$ matrix elements. These had odd spin $J=1,3,5,7,9$.
What emerged is shown in columns 3 and 4. The degeneracy is removed.
We have a $T=2$ state in the third column shifted up by 3~MeV and
in the fourth column a $T=0$ state unshifted. The wave function components
are different from what they are in the first two columns. The $T=2$
state is the unique state we were talking about---one that emerges
with any interaction, e.g. CCGI or delta. It is the double analog
of a state of four identical proton holes ($^{96}$Pd). The untangled
T=0 state in the 4'th column has vanishing {[}0,4{]} and{[}4,0{]}
components. Unlike the wave function in the 3'rd column this wave
function does not appear as an eigenstate for most other interactions.

In the last column, we list the other $T=2$, $v=4$ state. One sees
this on the list when one uses a seniority-conserving interaction
such as a delta interacton. However, for a general interaction, it
does not appear. This is because it gets mixed with the $T=2$, $v=2$
state. Only the state in the third column remains unscathed when we
turn on some arbitrary interaction---and only it does not end up being
degenerate with some other state.

There is also a unique v=4 J=6$^{+}$ state. With the pairing interaction
this is degenrate with another v=4 J=6$^{+}$ state and so the uniqueness
gets obscured. However with the Q.Q interaction, unlike the case for
J=4$^{+}$, this unique v=4 T=2 J=6$^{+}$state is not degenerate
with another state.Hence , even with Q.Q this state will appear in
a calculation.

There are other examples of confusions. The electric dipole moment
of the neutron would vanish if parity conservation holds. But at a
more important level, it vanishes if time reversal invariance holds.

\end{document}